\documentclass[twocolumn]{aastex701}
\usepackage[table]{xcolor}
\usepackage{hyperref}
\usepackage{amsmath}
\usepackage{amssymb}
\usepackage{float}
\usepackage{subcaption}
\usepackage{graphicx}
\usepackage{placeins}
\usepackage{multirow}
\usepackage{csquotes}
\usepackage{todonotes}
\usepackage{tikz}
\usetikzlibrary{arrows.meta, positioning, fit, calc}

\newcommand{\vect}[1]{\mathbf{\boldsymbol{#1}}}

\newcommand{\chieff}{\chi_\text{eff}}
\newcommand{\modot}{M_\odot}

\begin{document}

\title{BBH-Genesis: Disentangling Binary Black Hole Formation Channels with GWTC-4}

\author[orcid=0009-0006-3960-9405,gname=Shaunak, sname='Padhyegurjar']{Shaunak Padhyegurjar}
\affiliation{Department of Astronomy and Astrophysics, Tata Institute of Fundamental Research, Homi Bhabha Road, Navy Nagar, Colaba, Mumbai 400005, India}
\email[show]{shaunak.padhyegurjar@tifr.res.in}  

\author[orcid=0000-0002-3373-5236,gname=Suvodip, sname='Mukherjee']{Suvodip Mukherjee} 
\affiliation{Department of Astronomy and Astrophysics, Tata Institute of Fundamental Research, Homi Bhabha Road, Navy Nagar, Colaba, Mumbai 400005, India}
\email[show]{suvodip@tifr.res.in}

\begin{abstract}
The detected population of binary black holes (BBHs) from the gravitational wave (GW) data has made it possible to decipher their formation and evolution history over cosmic time. The complexity of astrophysical modeling of binary mergers makes it challenging to predict key signatures for different formation channels. As a result, one of the major avenues to discover the presence of different channels from detected GW events is through a \textit{data-driven way} which can isolate different scenarios. In this spirit, we developed a new inference pipeline \texttt{BBH-Genesis}  and applied it on the fourth GW catalog (GWTC-4) to identify the presence of multiple underlying distinct populations.  We find that the current population of all the binary events in GWTC-4 can be explained with the strongest evidence for only a two-channel scenario, hinting at the presence of a non-isolated binary formation channel. This sub-population can be further divided into a third channel with mild support towards formation in AGN exhibiting a slightly different effective spin and mass ratio correlation. In the future, with the detection of more events, it will be clearer whether it is necessary to consider at least three channels to explain the BBH events detected using GW observations.
\end{abstract}

\section{Introduction}\label{sec:intro}

Gravitational wave (GW) observations of binary black hole (BBH) mergers provide a unique window into the astrophysical processes that govern the formation and evolution of these systems. A central question in GW astrophysics is what formation channels give rise to the observed population of BBHs. The processes by which BBHs assemble and merge leave distinct imprints on the observable properties of BBHs which allow us to identify subpopulations of BBHs sharing similar properties corresponding to distinct formation channels.

Theoretically, BBH formation channels can be classified into two categories - isolated binary evolution in galactic fields and dynamical assembly in dense environments \citep{Mapelli:2020vfa, Gerosa:2021mno, Mandel:2021smh}. In the isolated channel, massive stellar binaries can undergo mass transfer and common envelope phases before collapsing into first generation (1G) black holes (BHs) that inspiral and eventually merge due to gravitational radiation. On the other hand, dense dynamical environments such as globular clusters, nuclear star clusters, and young star clusters assemble BBHs dynamically through processes like few-body encounters and exchange interactions \citep{Rodriguez:2015oxa}. Additionally, accretion disks of active galactic nuclei (AGN) are proposed to form migration traps where BHs could accumulate, assemble into BBHs and merge while interacting with the gas in the accretion disk \citep{2012MNRAS.425..460M, Mckernan:2017ssq, Yang:2019cbr, Tagawa:2019osr}. Such dynamical environments constitute the dynamical formation channels. 

These formation channels are expected to produce qualitatively different BBH populations. Isolated BBHs are typically expected to have low spin magnitudes with preferential alignment with the orbital angular momentum \citep{Rodriguez:2016vmx, 2019MNRAS.485.3661F, Ma:2019cpr, Fuller:2019sxi} and moderate to near-unity mass ratios \citep{Broekgaarden:2022nst, vanSon:2020zbk, Mandel:2015qlu}, while stellar evolution constraints imply that their component masses should largely lie below the pair-instability mass gap $(\sim45 \modot-120\modot)$ \citep{Belczynski:2016jno, Woosley:2021xba, Farmer:2019jed, Mapelli:2020vfa, Gerosa:2021mno, 2021ApJ...913...42W,Li:2023yyt}.

\begin{figure*}
    \centering
    \begin{minipage}{\textwidth}
        \centering
        \includegraphics[scale=1.]{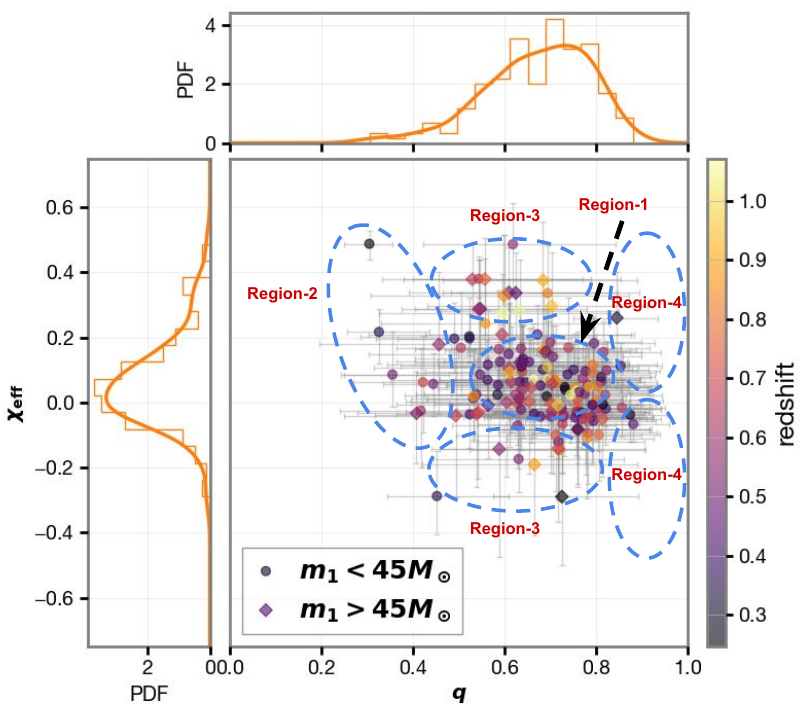}
    \end{minipage}
    \caption{Distribution of $(q, \chieff)$ for the events observed during GWTC4, including the two events GW241011\_233834 and GW241110\_124123 observed during O4b. Marker color represents redshift.}\label{fig:corr}
\end{figure*}

In dense star clusters, all BBHs are expected to have isotropic spin orientations due to the random nature of dynamical interactions \citep{Rodriguez:2016vmx}, whereas AGN disks can induce preferential (anti-)alignment due to interactions with the surrounding gas \citep{Mckernan:2017ssq, Tagawa:2019osr}. Dynamical channels can also produce higher generation ($n$G; $n>1$) BHs that are remnants of previous BBH mergers. A fraction of remnants can be retained in the environment that subsequently tend to pair with lighter, more abundant lower generation BHs resulting in hierarchical mergers that are characterized by high spin magnitudes ($\chi\sim0.7$) inherited from the progenitor BBH and unequal mass ratios $(q\approx0.5)$, largely independent of the formation channel \citep{Fishbach:2017dwv, Gerosa:2017kvu, Rodriguez:2019huv, Gerosa:2021mno}. Such hierarchical mergers can preferentially populate the pair-instability mass gap. First-generation (1G+1G) mergers still dominate the overall dynamical population and are expected to exhibit near-equal mass ratios $(q\approx1)$ as a consequence of mass segregation and dynamical exchange interactions.

{There exist significant theoretical uncertainties in stellar evolution modeling (such as common envelope evolution, supernovae natal kicks, tidal effects, stellar winds and dependence on parent star metalicity) and the astrophysical treatment of host environments in which BBHs form and merge, making precise first-principles predictions of observable merger rates of different channels challenging \citep{Mapelli:2020vfa, Mapelli:2021taw, Mandel:2025ngw}. However, with the increasing size of the gravitational wave transient catalog, it is becoming possible to distinguish and infer the relative contributions of different channels within the observed BBH population by modeling the robust features described above.}

The size of the latest GWTC-4 dataset \citep{LIGOScientific:2025slb} has enabled detailed population studies, with some studies providing strong evidence for the existence of a subpopulation of hierarchical mergers, possibly (but not limited to) of star cluster origin \citep{Antonini:2024het, Antonini:2025ilj, Tong:2025xir, Vijaykumar:2026zjy, Plunkett:2026pxt}, suggesting the existence of more than one formation channel. A few studies also hint toward a subpopulation consistent with the AGN channel \citep{Li:2025iux, Bartos:2026xlt}. See also \citep{Ray:2026uur} who hint towards presence of three channels and \citep{Afroz:2024fzp,Afroz:2025ikg} who hint towards multiple channels using a phase space approach.

With more than 150 GW events detected by the LIGO-Virgo-KAGRA Collaboration \citep{LIGOScientific:2014pky, VIRGO:2014yos, Virgo:2019juy, Virgo:2022ysc, KAGRA:2013rdx, Aso:2013eba, KAGRA:2020tym}, the key astrophysical property which can be inferred without performing any population-model-dependent search is the existence of correlation between different physical parameters which are inferred from GW data. In Fig. \ref{fig:corr}, we depict the correlation between the median value of the mass ratio $q \equiv m_2/m_1$ (where the component mass $m_2$ is less than $m_1$) and effective spin parameter $\chieff= (m_1\vec{\chi}_1.\vect{L}_{\rm orb}+m_2\vec{\chi}_2.\vect{L}_{\rm orb})/(m_1+m_2)$ for all 155 BBH events considered in this analysis (see Section below for our sample choice), along with their redshifts (shown in colorbar) and $68\%$
C.I of their posterior distribution \citep{gwtc4v2, gwtc3v2, gwtc2v2, gwtc4+}. The figure shows a few interesting features: 
\begin{itemize}
    \item \textbf{Region-1}: There exists a clustering between mass ratio and effective spin around 0.7 and 0 respectively, with no strong clustering in their redshift values. 
    \item \textbf{Region-2}: There exists a population of events which exhibit $|\chieff|>0$ and tend to be more at low redshift and low mass ratios ($q\lesssim0.4$).
    \item \textbf{Region-3}: There is a clustering in effective spin around $|\chieff|=0.3$ towards higher redshift values at a mass ratio around 0.5--0.7. 
    \item \textbf{Region-4}: There exists a lack of population of  $|\chieff|>0$ and mass ratio higher (close to one) in the existing data set. 
\end{itemize}
Certainly, these observations are based on visual inspection of the properties of the detected sample of events and are impacted by GW selection effects, but the evident correlation can be effectively used to disentangle underlying subpopulations. In this work, we developed a hierarchical Bayesian inference code \texttt{BBH-Genesis}  which can perform parametric data-driven inference of multiple population channels contributing to the observed GW events and allows to make a robust identification of them based on well determined correlations between the astrophysical parameters measured from GW strain data. 

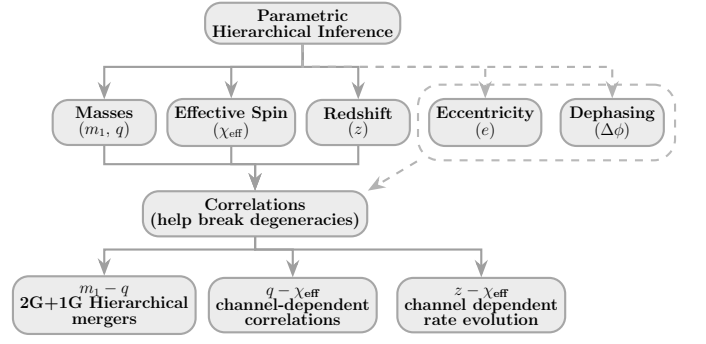
\begin{figure}
\centering
\begin{tikzpicture}[scale=0.525, every node/.style={transform shape}, 
  solidbox/.style={
    draw=gray!70, thick,
    rounded corners=6pt,
    fill=gray!15,
    font=\bfseries,
    minimum width=2.6cm,
    minimum height=0.9cm,
    inner sep=5pt,
    align=center
  },
  dashbox/.style={
    draw=gray!60, thick, dashed,
    rounded corners=6pt,
    fill=white,
    font=\bfseries,
    minimum width=2.6cm,
    minimum height=0.9cm,
    inner sep=5pt,
    align=center
  },
  arrow/.style={
    ->, >=Stealth, thick, color=gray!75
  },
  dasharrow/.style={
    ->, >=Stealth, thick, color=gray!60, dashed
  },
]
 
\node[solidbox, minimum width=3.8cm] (PHI) at (0, 0)
  {\large Parametric\\ \large Hierarchical Inference};
 
\node[solidbox] (masses) at (-5.0, -2.5) {\large Masses\\ \large $(m_1,\,q)$};
\node[solidbox] (spin) at (-1.8, -2.5) {\large Effective Spin\\ \large $(\chi_\mathrm{eff})$};
\node[solidbox] (redshift) at ( 1.4, -2.5) {\large Redshift\\ \large $(z)$};
\node[solidbox] (ecc) at ( 4.6, -2.5) {\large Eccentricity\\ \large $(e)$};
\node[solidbox] (deph) at ( 7.8, -2.5) {\large Dephasing\\ \large $(\Delta\phi)$};
 
\node[draw=gray!55, dashed, thick, rounded corners=8pt,
      fit=(ecc)(deph), inner sep=4pt, minimum height=1cm] (dashregion) {};

\node[solidbox, minimum width=4.4cm] (corr) at (-1.2, -4.8)
  {\large Correlations\\ \large (help break degeneracies)};

\node[solidbox, minimum width=3.0cm] (m1q)  at (-5.0, -7.1)
  {\large {$m_1-q$}\\ \large 2G+1G Hierarchical\\ \large mergers};
\node[solidbox, minimum width=3.0cm] (qchi) at (-0.25, -7.1)
  {\large {$q-\chieff$}\\ \large channel-dependent\\ \large correlations};
\node[solidbox, minimum width=3.0cm] (zchi) at ( 4.5, -7.1)
  {\large {$z-\chieff$}\\ \large channel dependent\\ \large rate evolution};
 
\def\busA{-1.1}
 
\draw[arrow] (PHI.south) -- (0,\busA) -| (masses.north);
\draw[arrow] (PHI.south) -- (0,\busA) -| (spin.north);
\draw[arrow] (PHI.south) -- (0,\busA) -| (redshift.north);
\draw[dasharrow](PHI.south) -- (0,\busA) -| (ecc.north);
\draw[dasharrow](PHI.south) -- (0,\busA) -| (deph.north);
 
\def\busB{-3.55}
 
\draw[arrow] (masses.south) -- (-5.0,\busB) -- (-1.2,\busB) -- (corr.north);
\draw[arrow] (spin.south) -- (-1.8,\busB) -- (-1.2,\busB) -- (corr.north);
\draw[arrow] (redshift.south) -- ( 1.4,\busB) -- (-1.2,\busB) -- (corr.north);
 
\def\busC{-5.7}
 
\draw[arrow] (corr.south) -- (-1.2,\busC) -| (m1q.north);
\draw[arrow] (corr.south) -- (-1.2,\busC) -| (qchi.north);
\draw[arrow] (corr.south) -- (-1.2,\busC) -| (zchi.north);

\draw[dasharrow] (dashregion.south west) -- (corr.north east);
 
\end{tikzpicture}
\caption{Landscape of GW population studies setup in \texttt{BBH-Genesis}.}
\label{fig:flowchart}
\end{figure}

We describe the basic flowchart of \texttt{BBH-Genesis}  in Fig. \ref{fig:flowchart}. It primarily relies on correlations in the data such as $(m_1, q)$ and $(q,\chieff)$, arising from the channel-specific features described above, to distinguish between the channels in the observed GW data. Currently it is limited to only mass $(m_1, q)$, effective spin $(\chieff)$ and redshift $(z)$ parameters, but could be easily extended to include parameters like eccentricity $(e)$ and orbital dephasing $(\Delta\phi)$ in the future to further help disentangle the channels. {The parametric data-driven search technique like \texttt{BBH-Genesis} differs from the phase-space technique \citep{Afroz:2024fzp, Afroz:2025efn, Afroz:2025ikg}. In the phase-space technique, one identifies the observational parameter space for different physical models, and can identify any outlier population of events, and can identify the probability for different events to be associated with different formation models. On the other hand, the analysis techniques such as \texttt{BBH-Genesis} can capture the global property of the entire population of the GW events present in the catalog in terms of different parametric models and can identify which parametric model is sufficient to explain most of the data, without making any judgment on outliers.}  

In application of the \texttt{BBH-Genesis} code on the latest GWTC-4 catalog, we consider two modeling scenarios: (i) \textbf{Two-channel model:} which searches for two distinct subpopulations in the data based on their mass ratio, effective spin, and merger rate distributions, and (ii) \textbf{Three-channel model:} which searches for three distinct subpopulations in the data based on their mass ratio, effective spin, and merger rate distributions, with a specific form of merger rate evolution for BBHs merging in AGN disks for one of the sub-population. Using a hierarchical Bayesian inference on the GWTC-4 catalog, we aim to address the following key questions:
\begin{itemize}
    \item Can the current data robustly determine the presence of more than one formation channels?
    \item Are there key correlations between GW source parameters in the GWTC-4 catalog that hint towards the astrophysical scenarios of different sub-populations? 
\end{itemize}

We find that while a two-channel model is supported by the data, current observations do not strongly favor resolving a third channel. However, the data shows interesting support toward a third channel consistent with the AGN formation scenario. The remainder of this paper is structured as follows. In Section \ref{sec:methods}, we describe the phenomenological models used in our study and provide astrophysical reasoning behind our modeling choices. In Section \ref{sec:results}, we describe the results of our population inference made using GWTC-4. We end with a summary and discuss the implications of our work and scope for future work in Section \ref{sec:discussion}. We provide a detailed description of our models in Appendix \ref{app:modeldetails}. All results are quoted at 90\% credible intervals, and luminosity distances are converted to redshift for the Planck best-fit cosmological model \citep{Planck:2015fie}.

\section{Methods} \label{sec:methods}

\subsection{Hierarchical Bayesian Framework}

We infer the properties of the observed population of BBHs using hierarchical Bayesian inference \citep{Loredo:2004nn, ThraneTalbot2019, Vitale:2020aaz}. The population distribution $\pi(\vect{\theta}|\vect{\Lambda})$ of source parameters $\vect{\theta}$ depends on model-dependent hyperparameters $\vect{\Lambda}$ that control the shape of the distribution resulting from different physical processes. The aim here is to estimate the hyperparameter posterior $p(\vect{\Lambda}|\vect{\mathcal{D}})$ using data of $N_\text{det}$ observed GW events $\vect{\mathcal{D}}=\{\vect{d}_1,\ldots,\vect{d}_{N_\text{det}}\}$. The (rate marginalized) population likelihood is given by \citep{LIGOScientific:2025pvj}
\begin{align}
    \mathcal{L}(\vect{\mathcal{D}}|\vect{\Lambda}) \propto \prod_{i=1}^{N_\text{det}}\frac{\int d\vect{\theta}\mathcal{L}(\vect{d}_i|\vect{\theta})\pi(\vect{\theta}|\vect{\Lambda})}{\xi(\vect{\Lambda})},\label{eqn:poplike}
\end{align}
where $\mathcal{L}(\vect{d}_i|\vect{\theta})$ are the likelihoods of individual events and $\xi(\vect{\Lambda})$ is the selection function given by
\begin{align}
    \xi(\vect{\Lambda}) = \int_{\rho(\vect{d})>\rho_\text{th}}d\vect{d}d\vect{\theta}\mathcal{L}(\vect{d}|\vect{\theta})\pi(\vect{\theta}|\vect{\Lambda}).\label{eqn:selfn}
\end{align}
The selection function corrects for the Malmquist bias \citep{1922MeLuF.100....1M} that arises from the selective efficiency of GW detectors in observing events that pass the detection threshold $\rho_\text{th}$ in the detection statistic $\rho(\vect{d})$ (SNR). 

The integrals in Eqs.\ \eqref{eqn:poplike} and \eqref{eqn:selfn} are practically intractable, so we estimate them using Monte Carlo sampling instead. The population likelihood can be estimated by reusing posterior samples of individual events obtained through parameter estimation as \citep{Tiwari:2017ndi, Farr2019, Essick:2022ojx}
\begin{align}
    \hat{\mathcal{L}}(\vect{d}_i|\vect{\Lambda}) \approx \frac{1}{N_\text{PE}}\sum_{j=1}^{N_\text{PE}}\frac{\pi(\theta_{ij}|\vect{\Lambda})}{\pi(\theta_{ij})}.
\end{align}
here $\pi(\theta_{ij}|\vect{\Lambda})$ is the event posterior, and $\pi(\theta_{ij})$ is the prior used for that event. The selection function can be estimated using a large number of \enquote{injections} drawn from a broad distribution and recovered (found) by event detection pipelines \citep{Essick:2025zed}. The selection function estimator is given as \citep{Tiwari:2017ndi, Farr2019, Essick:2022ojx}
\begin{align}
    \hat{\xi}(\vect{\Lambda}) \approx \frac{1}{N_\text{draw}} \sum_{j=1}^{N_\text{det}^\text{inj}} \frac{\pi(\vect{\theta}_j|\vect{\Lambda})}{\pi(\vect{\theta}_j|\vect{\Lambda}_\text{draw})}.
\end{align}

We model the BBH population as a mixture of subpopulations corresponding to distinct formation channels. The joint distribution of source parameters $\vect{\theta} = (m_1, q, \chieff, z)$ is written as
\begin{align}
    \pi(m_1,  &q, \chieff, z|\vect{\Lambda}) \nonumber \\
    \phantom{\mathrel{=}} &= \pi(m_1|\vect{\Lambda}_m)\Biggl[\, \sum_i f_i\pi_i(q, \chieff,  z|\vect{\Lambda}_i) \Biggr], \label{eqn:popmodel}
\end{align}
where $\{f_i\}$ are mixing fractions of the channels satisfying $\sum f_i = 1$, $\vect{\Lambda}_m$ are the hyperparameters of the primary mass distribution and $\vect{\Lambda}_i$ are channel specific hyperparameters. We use the \textsc{Power Law + Two Peaks} model for $\pi(m_1)$, common to all channels. Recent works of \citep{Ray:2026uur} also included features in the primary mass distribution to disentangle formation channels, however, there remains significant uncertainty in the knowledge of mass distributions of different channels. Moreover, hints of dynamically formed BBHs have been found across the mass spectrum \citep{Tong:2025xir, Banagiri:2025dmy,Plunkett:2026pxt, Ray:2026uur, Bartos:2026xlt}, which makes reliable modeling of channel-specific primary mass distribution difficult, thus motivating our choice of a common primary mass distribution. Instead, we explore distinguishing features in $(q, \chieff, z)$ to disentangle subpopulations of formation channels.

For each channel, we factorize the distribution as
\begin{align}
    \pi_i(q, \chieff, z|\vect{\Lambda}_i) = \pi_i(q|m_1, \vect{\Lambda}_i) \pi_i(\chieff|q, \vect{\Lambda}_i)\pi_i(z|\vect{\Lambda}_i). \label{eqn:chmodel}
\end{align}
The redshift distribution for each channel has the general form
\begin{align}
    \pi_i(z|\vect{\Lambda}_i) = \left(\frac{dV_c}{dz}\frac{1}{1+z} \right) R_0\psi_i(z|\vect{\Lambda}_i),\label{eqn:chredshift}
\end{align}
where $R_0$ is the total local merger rate (redshift $z=0$) and $\psi(z = 0|\vect{\Lambda}) = 1$. The construction of our model naturally gives the local merger rates for each channel as $R_{0,i} = f_iR_0$. We describe below the modeling choices for our two-channel and three-channel models with model variations summarized in Table \ref{tab:models}.

\begin{table*}
    \begin{tabular}{|c|c|c|c|c|}
       \hline
        Model & Channel & $\pi_i(q|m_1,\vect{\Lambda}_i)$ & $\pi_i(\chieff|q,\vect{\Lambda}_i)$ & $R(z)$ \\
        \hline
        \multirow{2}{*}{I} & 1 & $\pi_1(q|m_1, \vect{\Lambda}_1)$ & $\pi_1(\chieff|q, \vect{\Lambda}_1)$ & $R_0(1+z)^{\kappa_1}$ \\
         & 2 & $\pi_2(q|m_1, \vect{\Lambda}_2)$ & $\pi_2(\chieff|q, \vect{\Lambda}_2)$ & $R_0(1+z)^{\kappa_2}$ \\
        \hline
        \multirow{2}{*}{II} & 1 & Same as I & $\pi_1(\chieff|q, \vect{\Lambda}_1)$ with $U(-0.47,0.47)$ & Same as I \\
         & 2 & Same as I & $\pi_2(\chieff|q, \vect{\Lambda}_2)$ with $U(-0.47,0.47)$ & Same as I \\
        \hline
        \multirow{2}{*}{III} & 1 & Same as I & Same as II & Same as I \\
         & 2 & $\pi_2(q|m_1, \vect{\Lambda}_2)$ for $qm_1<m_\text{gap}$ & Same as II & Same as I \\
        \hline
        \multirow{3}{*}{IV} & 1 & Same as I & Same as II & $R_0\psi_1^\text{td}(z)$ \\
         & 2 & Same as I & Same as II & Same as I \\
          & 3 & $\pi_3(q|m_1,\vect{\Lambda}_3)$ & $\pi_3(\chieff|q, \vect{\Lambda}_3)$ with $U(-0.47,0.47)$ & $R_3^\text{AGN}(z)$ \\
        \hline
        \multirow{3}{*}{V} & 1 & Same as I & Same as II & Same as IV \\
         & 2 & $\pi_2(q|m_1, \vect{\Lambda}_2)$ for $qm_1<m_\text{gap}$ & Same as II & Same as I \\
          & 3 & $\pi_3(q|m_1, \vect{\Lambda}_3)$ for $qm_1<m_\text{gap}$ & Same as IV & Same as IV \\
        \hline
    \end{tabular}
    \caption{Summary of models studied in this work.}
    \label{tab:models}
\end{table*}

\subsection{Two-channel model}

In the two-channel model, the population is decomposed into Channel-1 (Ch-1; $i=1$) and Channel-2 (Ch-2; $i=2$) capturing key features described above. The mass ratio distribution for Channel-1 is modeled as
\begin{align}
    \pi_1(q|m_1, \vect{\Lambda}_1) \propto \begin{cases}
        q^{\beta_1} & \quad m_1 < m_\text{gap} \\
        0 & \quad \text{otherwise}
    \end{cases}, \label{eqn:qiso}
\end{align}
where the mass threshold $m_\text{gap}$ is imposed on primary mass. This condition imposes both components to lie below the mass gap as expected in the isolated channel. For Channel-2, we use
\begin{align}
    \pi_2(q|m_1, \vect{\Lambda}_2) \propto \xi_2 q^{\beta_2} + (1-\xi_2)\mathcal{N}_{[0, 1]}(\mu_{q,2},\sigma_{q,2}) \label{eqn:qnoniso}
\end{align}
as the default model. The Gaussian component seeks to capture the overdensity of events near $q\approx0.5$ in Region-3. We also impose the condition $qm_1 < m_\text{gap}$ in Models III, to allow the primary to exceed the mass gap while enforcing the secondary to remain below it, since we expect the dynamical channels to produce hierarchical mergers mainly of the 2G+1G kind. Low mass smoothing is applied to the above mass ratio models using a tapering function with independent parameters for $m_2$ than in the primary mass distribution.

The $\chieff$ distribution is modeled as a mixture between a Gaussian and a uniform distribution, with the fraction of the Gaussian component depending on $q$ as a power law $Aq^n$:
\begin{align}
    \pi_i(\chieff|q,\vect{\Lambda}_i) = &A_iq^{n_i}\mathcal{N}(\mu_i^{\chieff}, \sigma_i^{\chieff})\nonumber \\
    \phantom{\mathrel{=}} &\qquad+ (1 - A_iq^{n_i})U(\chi^\text{min}_i, \chi^\text{max}_i). \label{eqn:chieffmixture}
\end{align}
In all models except Model-I, we fix the uniform component to $U(-0.47, 0.47)$ . This is motivated by the fact that isolated BBHs have low, aligned spins, whereas BBHs in dynamical channels have isotropic spin orientations, with hierarchical mergers having high spins. We do not impose either component on a channel, but infer the mixing fractions instead. The $q$-dependence of the mixing fractions allows us to capture the $q-\chieff$ correlation. We model the redshift distribution of merger rate for both channels using separate power laws
\begin{align}
    \psi_i(z) = (1+z)^{\kappa_i}.
\end{align}

\subsection{Three-channel model}
 
In the three-channel model, the population is decomposed into three channels, namely Channel-1 (Ch-3; $i=1$), Channel-2 (Ch-2; $i=2$), and Channel-3 (Ch-3; $i=3$). This model tests whether the current data supports a third channel.

The mass ratio distribution for Channel-1 is the same as before, while Channels 2 \& 3 adopt the same functional form as in Eq.\ \eqref{eqn:qnoniso}, albeit with independent hyperparameters. in Model V, we also impose a condition that $qm_1< m_{\rm gap}$ (see Table \ref{tab:models}). The effective spin distributions for the three channels also follow the same functional form as in Eq.\ \eqref{eqn:chieffmixture} with independent hyperparameters, with the exception that the uniform component now has fixed width $U(-0.47, 0.47)$.

We modeled the Channel-1 redshift distribution using a power law time delay distribution convolved with the Madau-Dickinson cosmic star formation rate, written as \citep{Dominik:2014yma, Mandel:2015qlu, Vitale:2018yhm, Fishbach:2021mhp, Mukherjee:2021rtw, Karathanasis:2022hrb, Karathanasis:2022rtr}
\begin{align}
    \psi_1^\text{td}(z) = \frac{\int_z^\infty p_t(t_d|d, t_d^\text{min}, t_d^\text{max}) R_\text{SFR}(z_f) \frac{dt}{dz_f}dz_f}{\int_0^\infty p_t(t_d|d, t_d^\text{min}, t_d^\text{max}) R_\text{SFR}(z_f) \frac{dt}{dz_f}dz_f}.
\end{align}
The time delay distribution is defined as 
\begin{align}
    p_t(t_d|d, t_d^\text{min}, t_d^\text{max}) \propto \begin{cases}
        (t_d)^{-d} & \quad t_d^\text{min} < t_d < t_d^\text{max} \\
        0 & \quad \text{otherwise}
    \end{cases},
\end{align}
where time delay is given by $t_d=t(z_m)-t(z_f)$; $t(z)$ being the age of the universe at redshift $z$, $z_f$ and $z_m$ being the redshifts of formation and merger respectively. We use the Madau-Dickinson model \citep{Madau:2014bja}
\begin{align}
    R_\text{SFR}(z) = 0.015\frac{(1+z)^{2.7}}{1 + \left( \frac{1+z}{2.9} \right)^{5.6}} \; \modot \text{Mpc}^{-3}\text{yr}^{-1}
\end{align}
for cosmic star formation rate evolution.

The Channel-2 redshift evolution of the merger rate is modeled using a power law. The Channel-3 redshift distribution is modeled to be consistent with the AGN disk channel, which is expected to depend on the distribution of AGNs across cosmic time and the properties of the AGN disks hosting BBHs. Following \citep{Yang:2020lhq}, we model the AGN merger rate redshift evolution as
\begin{align}
    R_3^\text{AGN}(z) = \int_{L_\text{min}}^{L_\text{max}}\phi_L(L, z)d\log L \int_{\lambda_1}^1 \Gamma(\dot{m})P(\lambda|L, z)d\lambda, \label{eqn:agnredshift}
\end{align}
where $\phi_L(L,z)$ is the bolometric AGN luminosity function, $\Gamma$ is the average BBH merger rate in AGNs, $\dot{m}$ is the accretion rate of the central SMBH, $\lambda$ is the Eddington ratio and $P(\lambda)$ is the corresponding PDF. See \cite{Yang:2020lhq} and references therein for detailed derivation. The range of integration over AGN luminosity $L$ is $[L_\text{min}=10^{41}\text{erg }\text{s}^{-1}, L_\text{max}=3.15\times10^{14}L_\odot$]. We do not vary the AGN redshift model parameters in the three-channel models, but rather impose the distribution on the data to bring out a subpopulation following the above distribution, if present.

\section{Results} \label{sec:results}
We perform hierarchical Bayesian inference on our models using the cumulative GWTC-4 data publicly released by the LVK Collaboration \citep{gwtc4v2, gwtc3v2, gwtc2v2}. Our sample of events includes 153 BBH merger events detected up to O4a with FAR $\leq1\,\text{yr}^{-1}$ as well as the two exceptional events - GW241011\_233834 and GW241110\_124123, detected during O4b \citep{LIGOScientific:2025brd}, giving us a sample of 155 BBH merger events. We use the \textsc{NrSur7dq4} \citep{Varma:2019csw} posterior samples for events in GWTC-4 when available to minimize the effects of waveform systematics on hierarchical population inference \citep{Das:2026glo}, else we use \textsc{Mixed} posterior samples in the same spirit. For the two exceptional events, standard PE posterior samples were used \citep{gwtc4+}. We correct for selection effects using the suite of injections publicly released alongside GWTC-4 \citep{sensInj}. We impose a threshold $\sigma^2_{\ln\mathcal{L}}<1$ on the log-likelihood variance to manage the bias resulting from using a finite number of posterior samples per event in Monte Carlo sums in likelihood estimation. We describe below all the results for all the models considered in this analysis. In table \ref{tab:q_spins_z_hyperpriors}, we have summarized the findings. 

\subsection{Two-channel Model}

\subsubsection{Model-I} We infer the fractions of Channel-1 and Channel-2 to be $53\%$ and $47\%$ using our Model-I. The subpopulation of Channel-2 is characterized by a steeper power law $(\beta_2=2.81^{+3.31}_{-3.84})$ component in mass ratio distribution than Channel-1 $(\beta_2=1.80^{+1.75}_{-1.22})$, suggesting preference for near-equal mass binaries $(q\approx1)$. The Gaussian peak in mass ratio in Channel-2 appears at $\mu_{q,2}=0.56^{+0.37}_{-0.40}$ with width $\sigma_{q,2} = 0.58^{+0.35}_{-0.37}$, which is largely consistent with the overdensity of events around $q\approx0.5$ found in Region-3.

\begin{figure*}
    \centering
    \begin{minipage}{0.32\textwidth}
        \centering
        \includegraphics[scale=0.35]{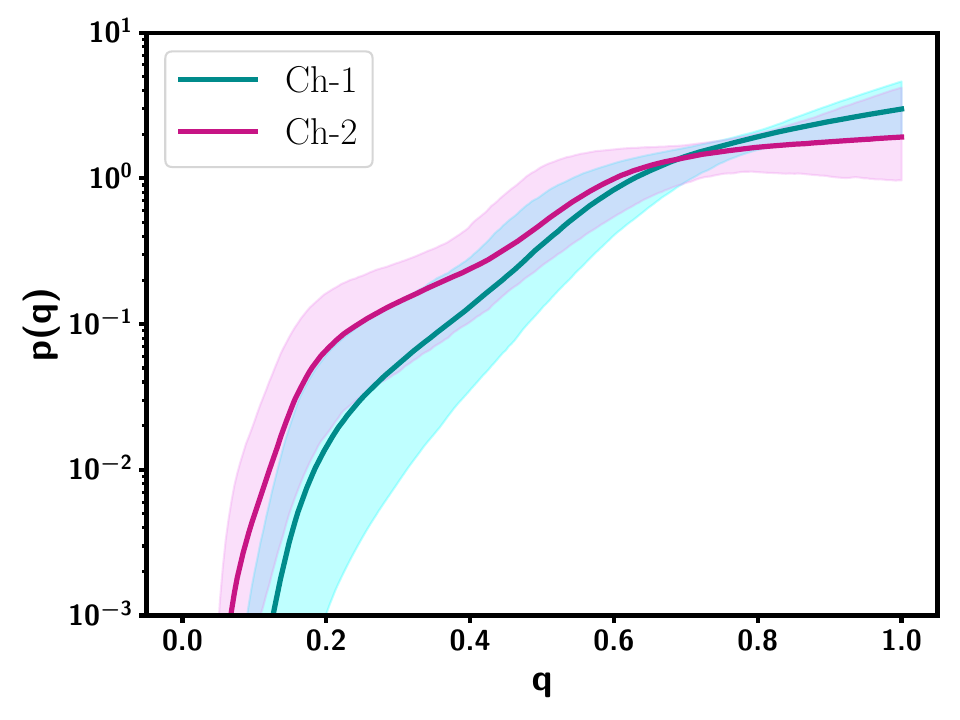}
    \end{minipage}
    \begin{minipage}{0.32\textwidth}
        \centering
        \includegraphics[scale=0.35]{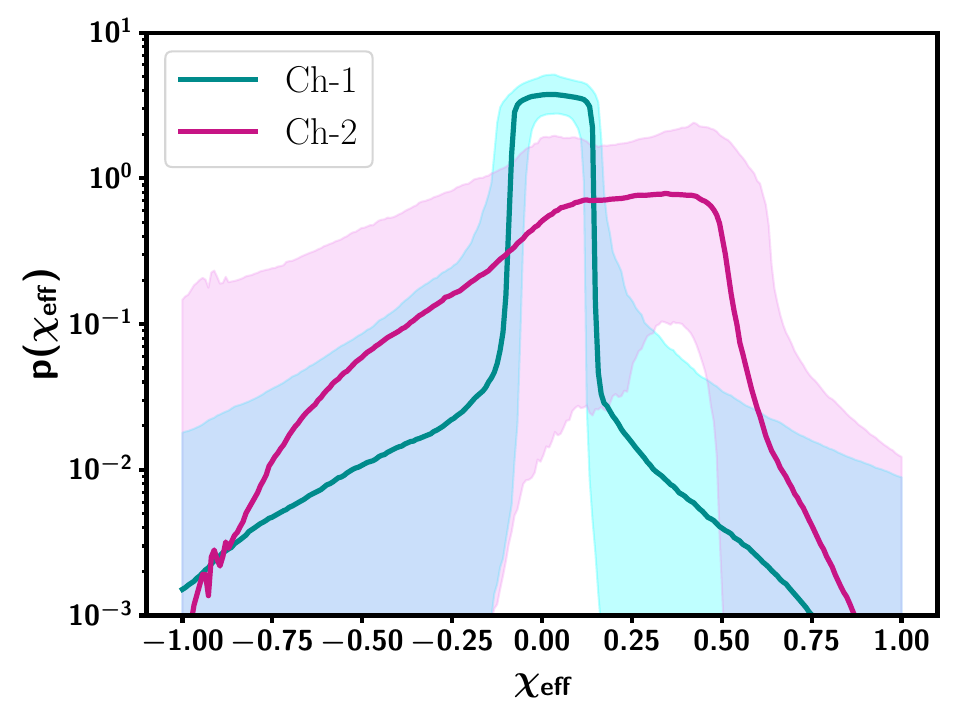}
    \end{minipage}
    \begin{minipage}{0.32\textwidth}
        \centering
        \includegraphics[scale=0.35]{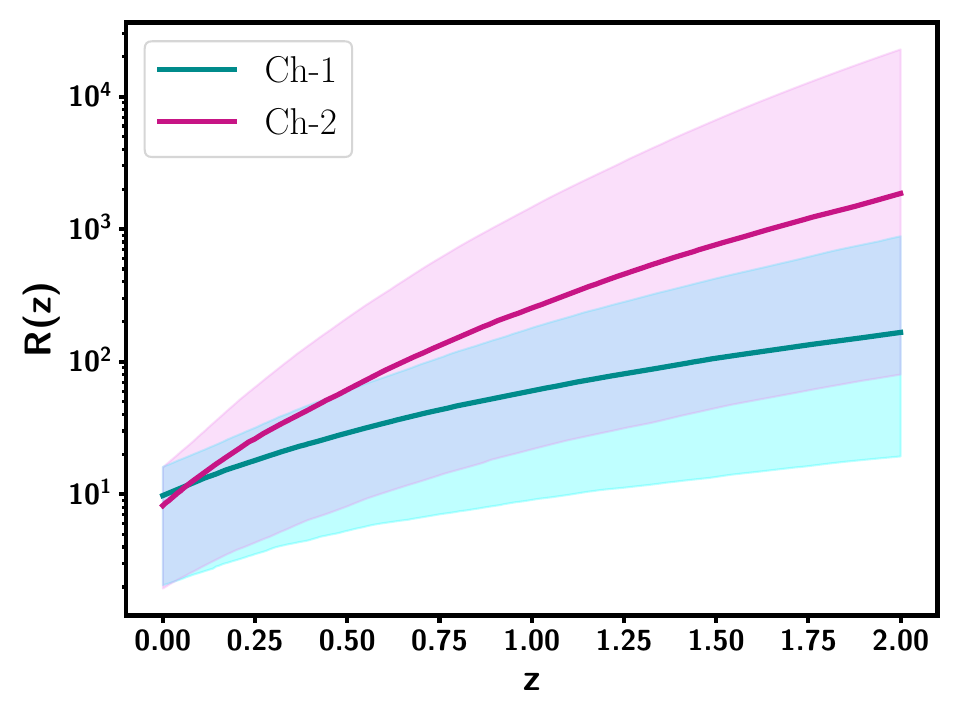}
    \end{minipage}
    \caption{Inferred distributions of mass ratio (\emph{left}), effective spin (\emph{center}), and redshift evolution of merger rate of channels (weighted by corresponding fractions) for the two-channel Model-I.}
    \label{fig:model1_q_chieff_z}
\end{figure*}

\begin{figure*}
    \centering
    \begin{minipage}{\textwidth}
        \centering
        \includegraphics[scale=.5]{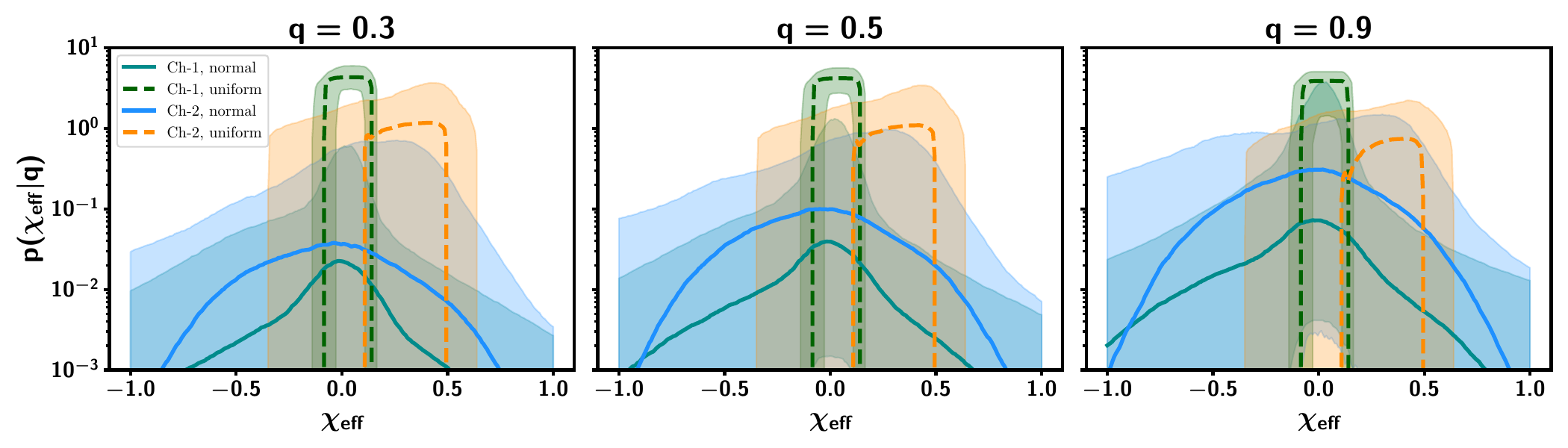}
    \end{minipage}
    \caption{Variation of normal and uniform components in effective spin distribution inferred using Model-I for different values of mass ratio.}
    \label{fig:model1_p_q_components}
\end{figure*}

\begin{figure}[h!]
    \centering
    \begin{minipage}{0.5\textwidth}
        \centering
        \includegraphics[scale=0.375]{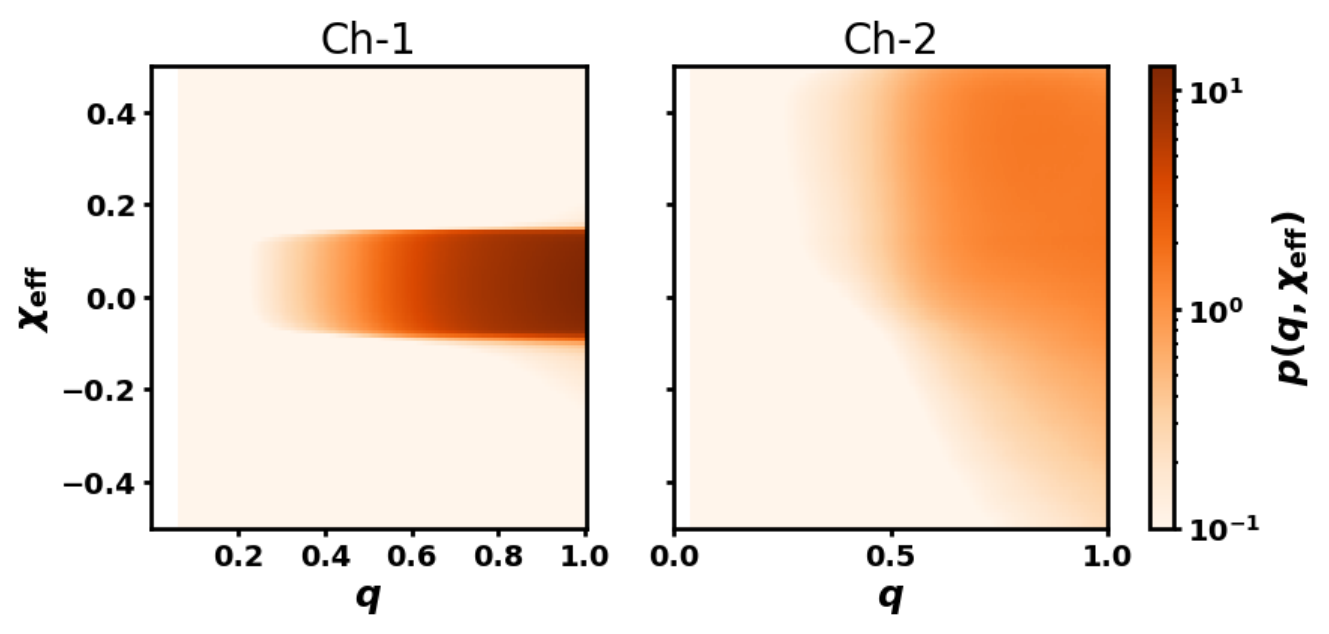}
    \end{minipage}
    \caption{Joint distribution $p(q,\chieff)$ inferred by our Model-I.}
    \label{fig:model1_p_q_2D}
\end{figure}

The transition mass scale $m_\text{gap}$ associated with the onset of the pair-instability mass gap\footnote{Note that the lower edge of the mass gap is not directly measured in GW population analyses. It is inferred as a transition scale in a model-dependent way.} is constrained to be $66.31^{+11.10}_{-11.98}\,\rm M_{\odot}$ when inferred only from transition in mass ratio distribution based on primary mass (Eq.\ \eqref{eqn:qiso}). This is higher than the previously reported estimate $\sim40-50\modot$ based on transition in effective spin distribution and imposing a cutoff in secondary mass distribution \citep{Karathanasis:2022rtr,Afroz:2025ikg,Antonini:2024het, Antonini:2025ilj, Tong:2025wpz, Tong:2025xir}. However, a transition mass scale inferred from effective spin distribution might not capture the PISN cutoff reliably since there is no sharp cutoff in the secondary mass distribution in $40-50\modot$, and the actual value could be higher \citep{Ray:2025xti}. Other works reporting a higher value for a  \citep{Xia:2026lwv}. See also \citep{Mukhamedzhanov:2026yha} who report a higher PISN cutoff from constraints on the S factor of $^{12}\text{C}(\alpha, \gamma)^{16}\text{O}$ reaction.

\begin{figure*}
    \centering
    \begin{minipage}{0.32\textwidth}
        \centering
        \includegraphics[scale=0.35]{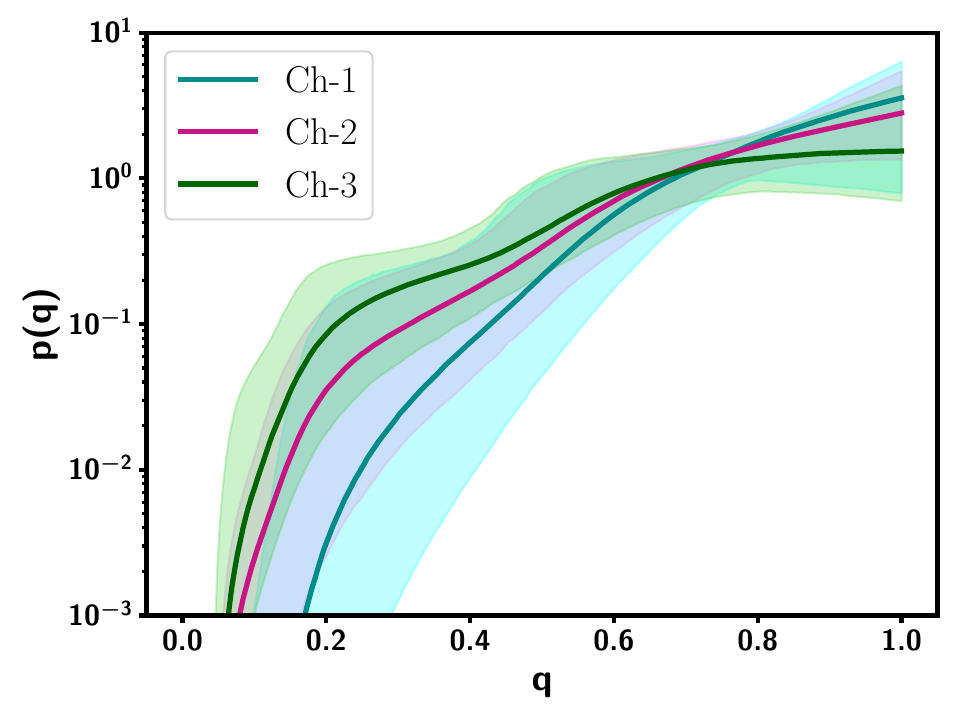}
    \end{minipage}
    \begin{minipage}{0.32\textwidth}
        \centering
        \includegraphics[scale=0.35]{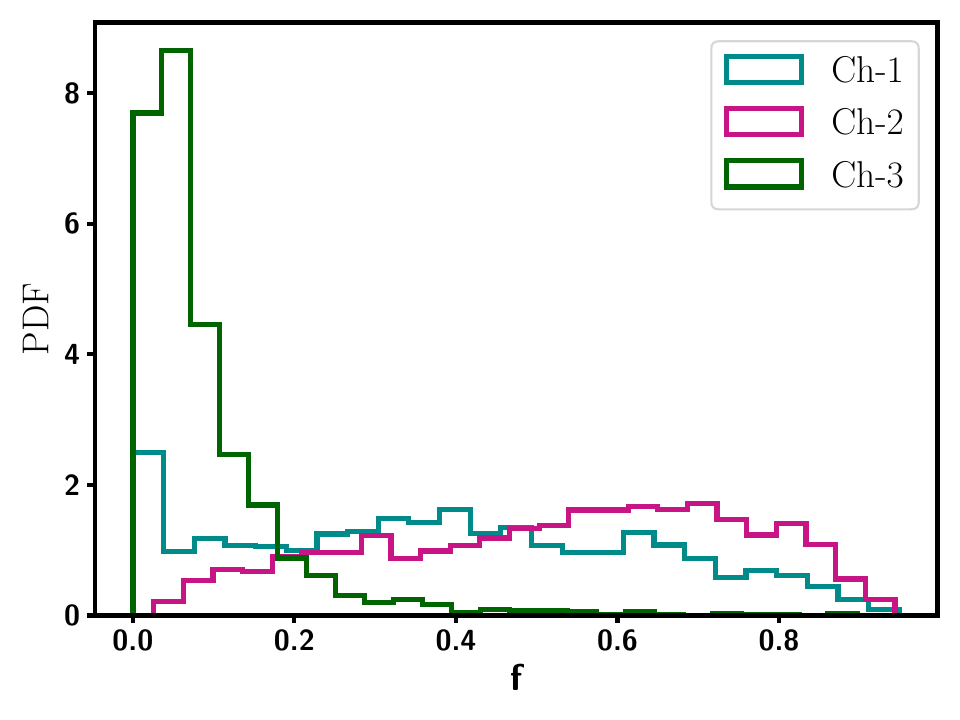}
    \end{minipage}
    \caption{Inferred mass ratio distributions (\emph{left}) and mixing fractions (\emph{right}) of the three channels in Model-V.}
    \label{fig:model5_q_f}
\end{figure*}

\begin{figure*}
    \centering
    \begin{minipage}{\textwidth}
        \centering
        \includegraphics[scale=0.475]{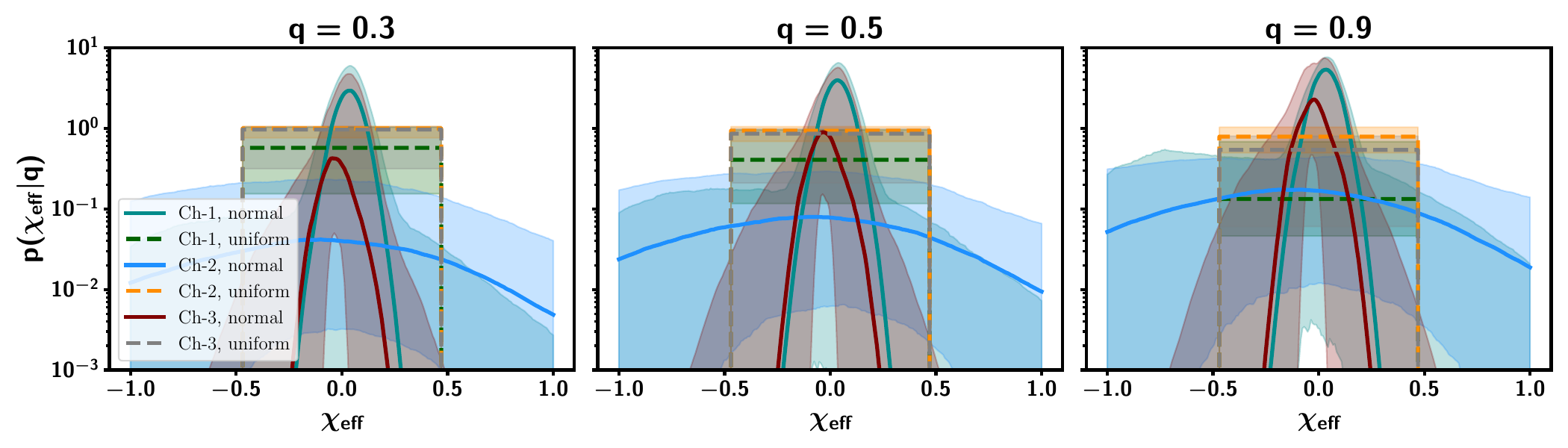}
    \end{minipage}
    \caption{Variation of normal and uniform components in effective spin distribution inferred using Model-I for different values of mass ratio.}
    \label{fig:model5_p_q_components}
\end{figure*}

\begin{figure*}
    \centering
    \begin{minipage}{\textwidth}
        \centering
        \includegraphics[scale=0.425]{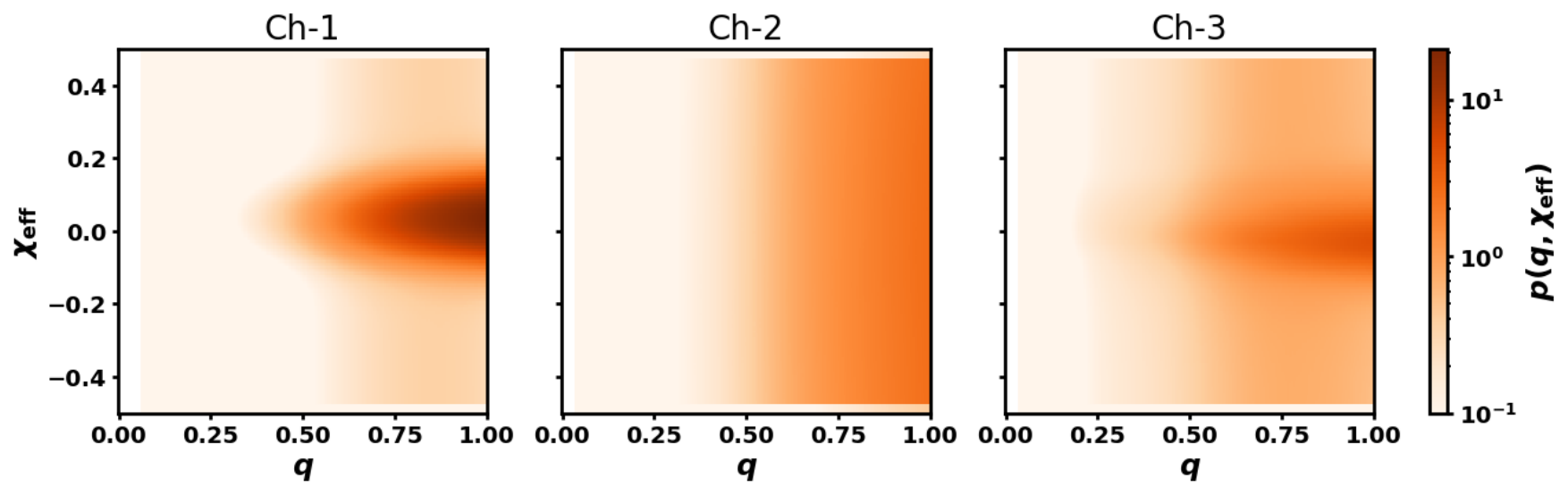}
    \end{minipage}
    \caption{Joint distribution $p(q, \chieff)$ inferred for Model-IV.}
    \label{fig:model5_q_chieff_2D}
\end{figure*}

The effective spin distribution provides a clear separation between channels. As seen in Fig.   \ref{fig:model1_q_chieff_z}, Channel-1 is characterized by a narrow distribution centered at $\chieff \approx 0$ consistent with Region-1 in Fig.   \ref{fig:corr} and isolated binary evolution, whereas Channel-2 favors a broad distribution centered near $\chieff\approx0.3$ capturing events with $|\chieff|>0$ in Region-3 consistent with dynamical evolution. Our model for $p(\chieff|q)$ also effectively captures the distinct $q-\chieff$ correlation for both channels, as shown in the variation of $p(\chieff|q)$ for different values of $q$ in Fig.   \ref{fig:model1_p_q_components}. The figure shows declining contribution of the Channel-2 uniform component as expected from the lack of events with $|\chieff|>0$ for $q\approx1$ in Region-4 in Fig.   \ref{fig:corr}, whereas the Channel-1 normal component increases slightly towards $q=1$, reducing the uniform component centered at $\chieff\approx0$. It is interesting to note that, however, given the flexibility of varying the endpoints of the uniform distribution in Model-I (see Eq.\ \eqref{eqn:chieffmixture}, the data favors suppression of the normal component in both channels and the uniform components alone capture the clustering of events at $\chieff\approx0$ and $|\chieff|>0$. This also explains the higher evidence measured for Model-I despite it having more parameters than the other two-channel models (see Table \ref{tab:q_spins_z_hyperpriors}).

The redshift evolution of merger rates of the two channels also shows a clear distinction with the power law slope of Channel-2 $(\kappa_2=4.96^{+1.81}_{-2.05})$ higher than Channel-1 $(\kappa_1=2.61^{+1.36}_{-1.00})$. Although Channel-2 appears to dominate over Channel-1 in the merger rate at higher redshifts, the higher power law index of Channel-2 can be explained by the presence of high-mass, high-redshift observations in Region-3 in Fig.   \ref{fig:corr}. These events preferentially occupy the region of parameter space associated with Channel-2 in our model and cause broadening of the $\chieff$ distribution with redshift as also observed by \citep{Biscoveanu:2022qac, Farah:2026jlc}.

\subsubsection{Model-II and Model-III}

The inferred fractions of Channel-1 and Channel-2 are $53\%$ and $47\%$ for Model-II and $58\%$ and $42\%$ respectively for Model-III, which are consistent with Model-I results. The mass ratio distributions for Models II and III also show similar trends as Model-I, with Channel-2 power laws steeper than Channel-1 and a Gaussian peak near $q\approx0.5$.

While the inferred value of $m_\text{gap}=65.36^{+12.39}_{-16.13}\modot$ for Model-II is consistent with Model-I, the same for Model-III, which also imposes the cutoff in secondary mass distribution, is even larger ($76.88^{+2.75}_{-4.23}\modot$). The posterior of $m_\text{gap}$ inferred from Model-III is also much narrower compared to Models I and II, as seen in Fig.   \ref{fig:corner_model2_model3}, which is also reflected in much smaller errors. This could be attributed to either the inference pushing the cutoff to a value where the $m_2$ distribution vanishes as a consequence of the PISN mass gap or an effect of prior choice. The estimate is quite close to the upper boundary of $80\modot$ of the chosen prior, and the inference could choose the largest possible values even if the underlying distribution isn't vanishing. A separate check is warranted for ruling out any prior effects. Additionally, the current data favors Model-II over Model-III with $\log_{10}BF_{\text{II}/\text{III}}=1.09$. While the rest of the parameters are consistent between Models II and III, this suggests that the data does not strongly support such a high transition mass scale, and that the inference is primarily driven by the treatment of the mass-gap cutoff rather than differences in the subpopulations.

With the increased PISN cutoff value, the power law index of Channel-2 $(\kappa_2=6.44^{+2.33}_{-2.23})$ also increased in Model-III, which can be again explained by the high-mass, high-redshift observations in Region-4.

\subsection{Three-channel Model}

We infer the fractions of Channels 1, 2, and 3 to be $38\%$, $55\%$, and $6\%$ respectively for Model-IV and $62\%$, $36\%$, and $2\%$ respectively for Model-V. Although the fractions are similar between the two models, the fractions of Channels 1 and 2 have flipped in Model-IV compared to Model-I. This behavior is likely a consequence of degeneracies between the phenomenological population components rather than a strong evidence for a genuine change of underlying astrophysical subpopulations.

The mass ratio distributions of Channels 1 and 2 are consistent with those in the two-channel models, with the Channel-2 power law being steeper than that for Channel-1 in both Model-IV and Model-V. However, the same for Channel-3 is comparable to Channel-1 for Model-V whereas it is much smaller $(\beta_3=0.38^{+5.57}_{-2.17})$ for Model-IV. The Gaussian peak in Channel-2 and Channel-3 appears at $q\approx0.6$ and $q\approx0.5$, respectively, for both three-channel models, consistent with the two-channel models. The estimate for $m_\text{gap}$ for Model-IV is consistent with Model-I and Model-II, while that of Model-V is consistent with Model-III, with the imposition of the cutoff in the secondary mass distribution making the difference in favoring Model-IV over Model-V, with $\log_{10}BF_{\text{IV/V}}=0.85$.

The effective spin distributions in Model-IV and Model-V share the same characteristics as the two-channel models, showing clear separation between the three channels in Figs.  \ref{fig:model5_p_q_components} and \ref{fig:model5_q_chieff_2D}. The $\chieff$-distribution of Channel-1 is well captured by a Gaussian centered at $\chieff\approx0$ whereas the uniform component dominates Channel-2 \& Channel-3. 

In the three-channel models, we adopted a realistic redshift model following cosmic star formation rate with a time delay for Channel-1 and a redshift model following the distribution of AGNs for Channel-3, capturing the AGN disk channel. We infer a minimum time delay of $0.86^{+0.90}_{-0.60}\,\text{Gyr}$ for Channel-1 using Model-IV whereas it increases to $1.14^{+0.99}_{-0.83}\,\text{Gyr}$ in Model-V. Importantly, we find $2\%$ and $6\%$ of the total BBH population following the fixed AGN redshift model in Eq.\ \eqref{eqn:agnredshift} for Models IV and V respectively (see Fig.\ \ref{fig:model5_q_f}), which is in line with the recent findings of \cite{Bartos:2026xlt} who found a $\sim10\%$ (90\% CI $[1\%,14\%]$) fraction of observed BBHs consistent with accretion-origin in AGN disks. A companion paper \citep{bbhgenesistimedelay} show the measurement of three sub-population of time-delay distribution based on their source properties.  

Three-channel models IV and V are disfavored over the two-channel model I substantially, with $\log_{10}BF_\text{IV/I}=-3.59$ and $\log_{10}BF_\text{V/I}=-4.44$. So, we find Model-I is the best-fit model for the data, and there is no strong support in the data to consider a three-channel model at present.  

\section{Discussion and Conclusion} \label{sec:discussion}
In the previous section, we have shown the results obtained using \texttt{BBH-Genesis}  for different scenarios, including two-channel and 
three-channel models from the GWTC-4 catalog. The results have found some of the key interesting features in the mass ratio and spin frame. The key messages from the analysis are as follows: 

\begin{itemize}
    \item The two-channel model shows a clear support towards the presence of two sub-population models exhibiting identifiable features with mass ratio, effective spin, and redshift.
    \begin{enumerate}
        \item The mass ratio distribution of the Channel-2 sub-population exhibits a broad peak around $q=0.6$. This feature can be identified with Region-3 identified in Fig. \ref{fig:corr}. This feature shows more support towards a uniform effective spin distribution for $\chieff>0$.
        \item The Channel-1 subpopulation shows more support towards a narrow Gaussian (narrow uniform in Model-I) peaked around $\chieff\approx0$ for higher values of mass ratio $q$ consistent with Region-1 in Fig.\ \ref{fig:corr}.
        \item The clustering of the events exhibiting a peak at a non-zero value of effective spin $\chieff$ shows more support toward high redshift (as denoted by Region-3), which is supported by this sub-population. We find that the sub-population which is showing a peak in $\chieff$ around 0.3, also supports a higher merger rate at higher redshift, in comparison to the other population. The fraction of both channels shows an equal preference and no clear dominance.
    \end{enumerate}
    \item The three-channel model, though it shows a weak Bayes factor in comparison to the two-channel model, has some interesting prominent features that stand out over the two-channel model.
    \begin{enumerate}
        \item One of the sub-population in the two-channel model gets segregated into a third sub-population showing a mild contribution of about $6\%$ towards a merger rate redshift distribution which follows the AGN redshift distribution. 
        \item This sub-population captures both the features in mass ratio distribution identified for the two-channel case (which matches with Region-1 and Region-3 identified in Fig. \ref{fig:corr}), but with a little stronger contribution.
        \item The features in the effective spin distribution associated with the Region-2 and Region-3 are supported by this third sub-population for the value of $\chieff \approx 0$ towards higher mass ratios, in contrast to lower mass ratios. 
    \end{enumerate}
\end{itemize}

These findings show that though the data can be explained by a three-population model, with about $6\%$ support towards a third channel, it is not favored over a two-channel model. However, the presence of an aligned spin component and misaligned spin components (with more support toward a higher merger rate at high redshift) is evident. Interestingly, our model captures the essential features in the mass and effective spin with redshift to identify some of the smoking gun features that can be associated with the dynamical channel over the isolated channel. The parametric data-driven approach considered in this analysis is able to classify the entire BBH events catalog into at least two sub-populations with disjoint features. A recent study has also pointed out the existence of three sub-populations \citep{Cheng:2026bpc}. 

The number of events used in the analysis is one of the major limiting factors in not being able to shed light on a potential third channel. However, with the help of upcoming observation data, the tentative evidence toward the third sub-population will be clearer. Another limitation from the modeling side remains in not incorporating other observables, such as eccentricity, dynamical friction effects, and recoil velocity, mainly due to poor measurements of these quantities with current GW detectors. However, improvements in these measurements in future will enable a richer exploration of channel-dependent features and help disentangle the subpopulations better. 
In the future, the \texttt{BBH-Genesis}  framework will be improved to capture these signatures. 

In conclusion, a data-driven parametric analysis method  \texttt{BBH-Genesis}  developed in this paper and applied to GWTC-4 has identified a few key features in the population of binary black holes detected by the LVK Collaboration and discovered the existence of at least two kinds of sub-populations and tentative hints towards the presence of a third sub-population. The decisive power in the method comes from the two segregated populations in the effective spin parameter distribution and its correlation with the mass ratio and redshift. This underlying identification of sub-population connects with the regions identified in the GWTC-4 event catalog. Our analysis discovers the potential sub-population in mass ratio--effective spin--redshift parameter space, which can be connected with two sub-class populations of isolated formation channel and dynamical formation channel. Moreover, we are able to identify support for the presence of BBH in AGN discs. However, such a model is not required to fit the catalog, as the Bayes factor favors a two-channel model more than a three-channel model. In summary, the detection of more GW sources from the new observation runs will enable the classification of sub-populations of BBH and shed light on their formation channels with higher precision in the coming years. 
 
\begin{acknowledgments}
The authors are grateful to Aniruddha Chakraborty for carefully reviewing the manuscript and providing useful suggestions to improve the draft. This work is a part of the ⟨Data$|$Theory⟩ Universe Lab, which is supported by the Department of Astronomy and Astrophysics at the Tata Institute of Fundamental Research (TIFR) and the Department of Atomic Energy, Government of India. We acknowledge the support of the Department of Atomic Energy, Government of India, under Project Identification No. RTI 4012. This research is supported by the Prime Minister Early Career Research Award, Anusandhan National Research Foundation, Government of India. We are also thankful for the computing resources provided by the ⟨Data$|$Theory⟩ Universe Lab. LIGO, funded by the U.S. National Science Foundation (NSF), and Virgo, supported by the French CNRS, Italian INFN, and Dutch Nikhef, along with contributions from Polish and Hungarian institutes. This collaborative effort is backed by the NSF’s LIGO Laboratory, a major facility fully funded by the National Science Foundation. 
\end{acknowledgments}

\begin{software}
    \ This work made use of the following software packages: \texttt{astropy} \citep{astropy:2013,astropy:2018,astropy:2022}, \texttt{Jupyter} \citep{2007CSE.....9c..21P,kluyver2016jupyter}, \texttt{matplotlib} \citep{Hunter:2007}, \texttt{numpy} \citep{numpy}, \texttt{pandas} \citep{mckinney-proc-scipy-2010,pandas-17229934}, \texttt{python} \citep{python}, \texttt{scipy} \citep{2020SciPy-NMeth,scipy-17467817}, \texttt{Bilby} \citep{bilby-paper,bilby-paper-2,Bilby-17533961}, \texttt{GetDist} \citep{Lewis:2019xzd,GetDist_20083735}, \texttt{Dynesty} \citep{Speagle:2019ivv}, \texttt{gwpopulation} \citep{Talbot2025}, \texttt{gwpopulation\_pipe} \citep{Talbot2021}.

    Software citation information aggregated using \texttt{\href{https://www.tomwagg.com/software-citation-station/}{The Software Citation Station}} \citep{software-citation-station-paper,software-citation-station-zenodo}.
\end{software}

\newpage

\appendix

\section{Additional plots and summary table} \label{app:modeldetails}

\begin{figure*}[h!]
    \centering
    \begin{minipage}{0.98\textwidth}
        \centering
        \includegraphics[scale=0.175]{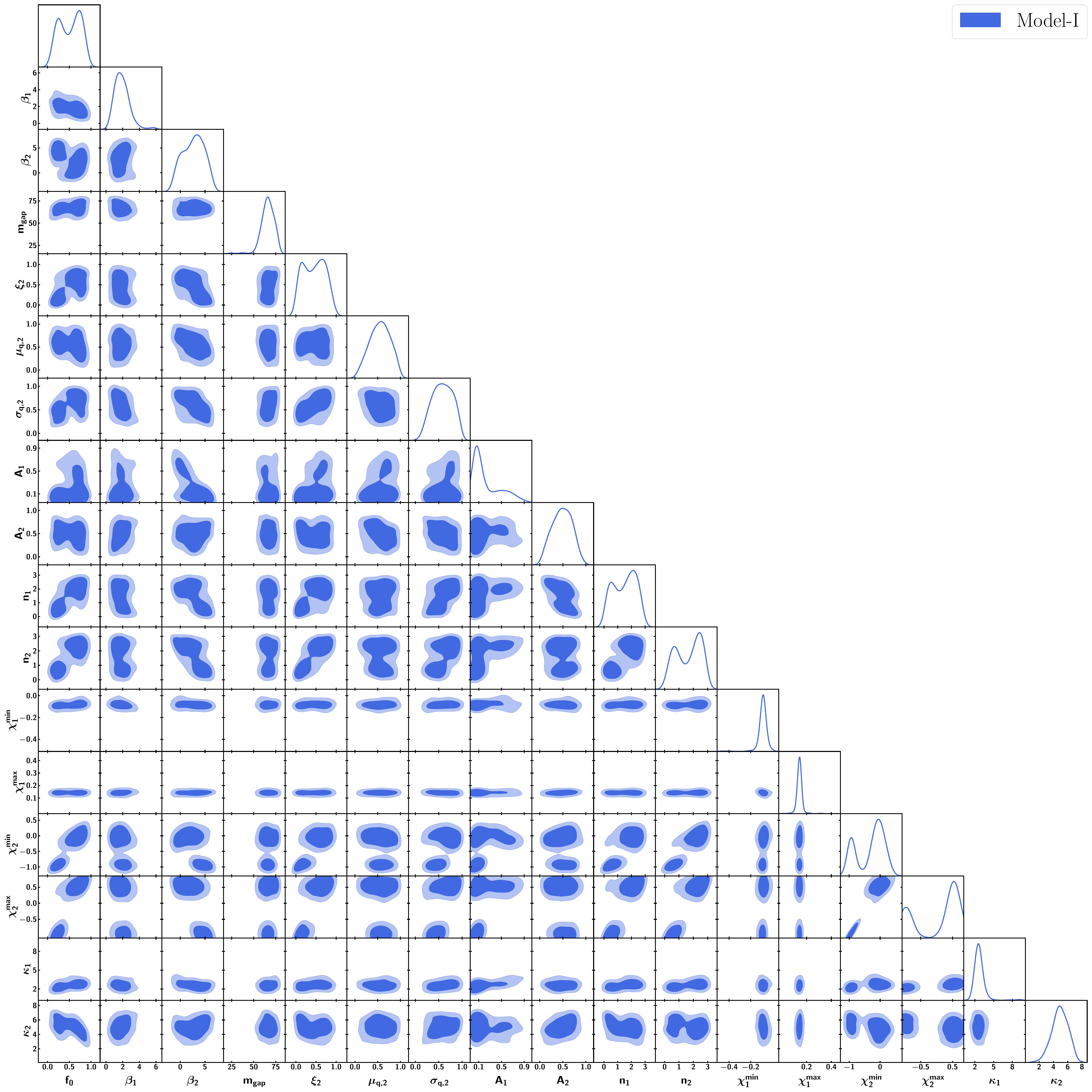}
    \end{minipage}
    \caption{Corner plot for select hyperparameters for Model-I.}
    \label{fig:corner_model1}
\end{figure*}

\begin{figure*}
    \centering
    \begin{minipage}{0.98\textwidth}
        \centering
        \includegraphics[scale=0.2]{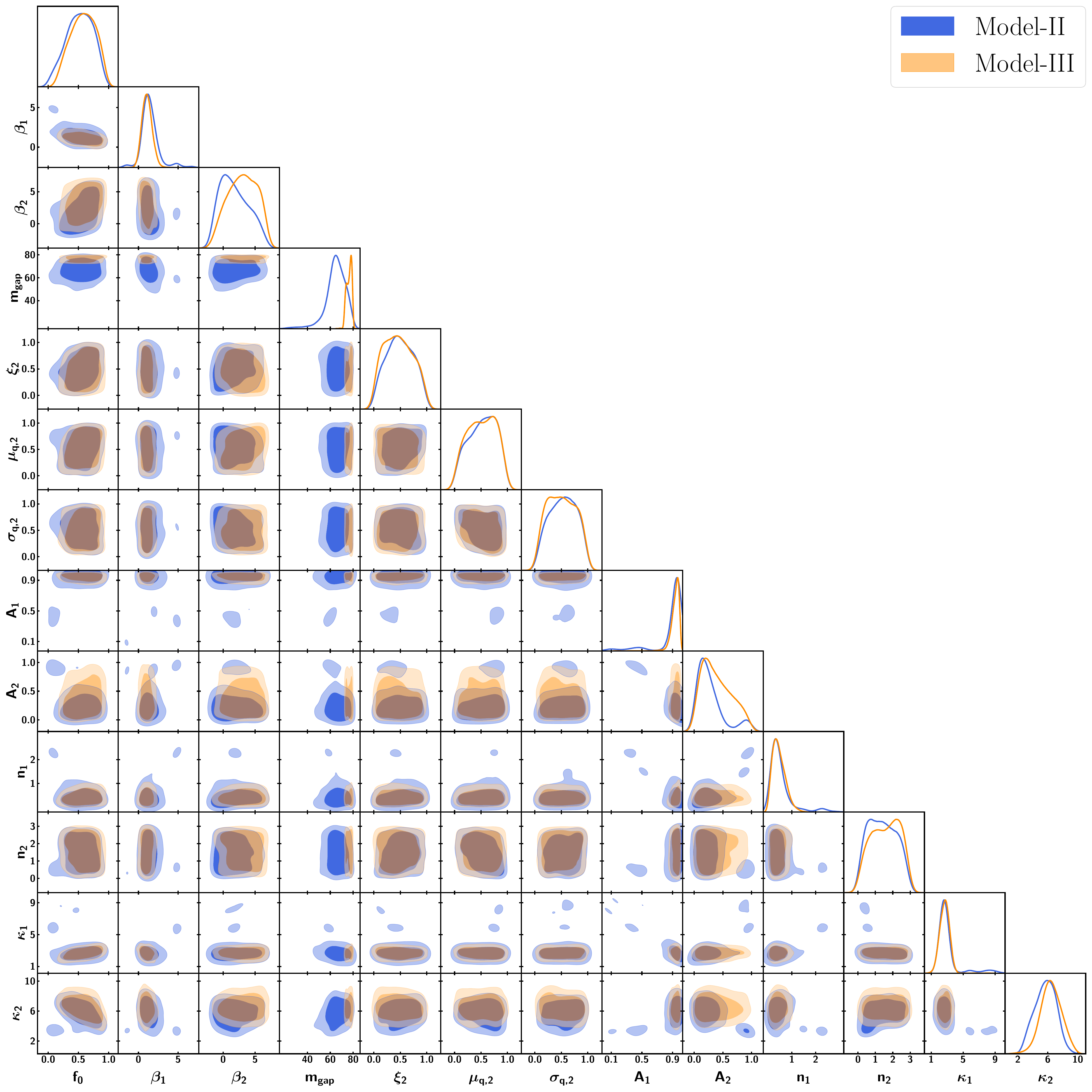}
    \end{minipage}
    \caption{Corner plot for select hyperparameters for Models II and III.}
    \label{fig:corner_model2_model3}
\end{figure*}

\begin{figure*}
    \centering
    \begin{minipage}{0.98\textwidth}
        \centering
        \includegraphics[scale=0.15]{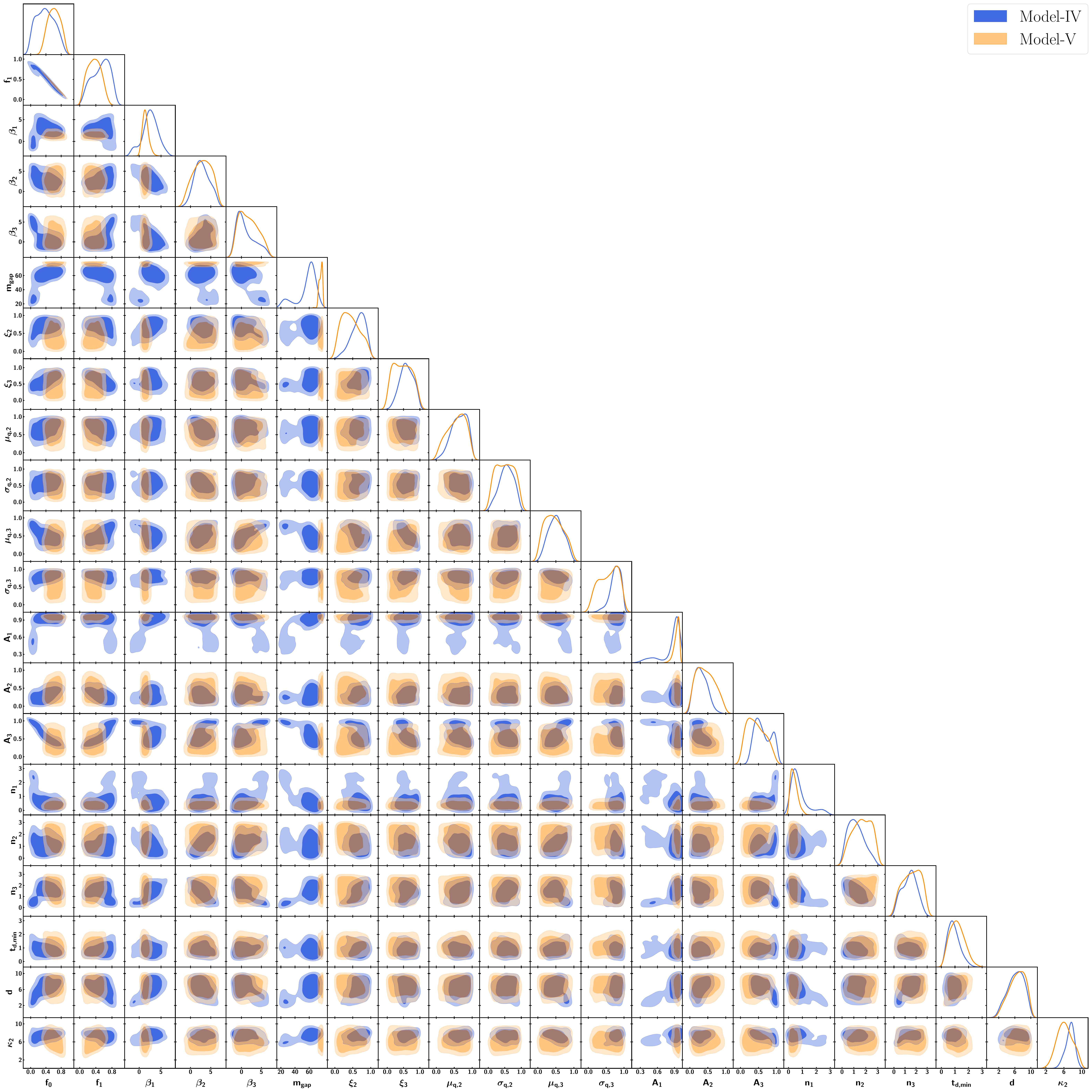}
    \end{minipage}
    \caption{Corner plot for select hyperparameters for Models IV and V.}
    \label{fig:corner_model4_model5}
\end{figure*}

Table \ref{tab:q_spins_z_hyperpriors} below provides constraints for all models considered in this work along with the choices of hyperparameters. Corner plots for select hyperparameters for all the models are shown in Figs.  \ref{fig:corner_model1}-\ref{fig:corner_model4_model5}.

\begin{table*}[h!]
    \centering
    \renewcommand{\arraystretch}{1.25}
    \begin{tabular}{c|c|ccccc}
        \hline\hline
        \multirow{2}{*}{\textbf{Parameter}} & \multirow{2}{*}{\textbf{Prior}} & \multicolumn{5}{c}{\textbf{Estimates (90\% CI)}} \\
         &  & Model-I & Model-II & Model-III & Model-IV & Model-V \\
        \hline
        $\log_{10}Z$ & - & $-1926.81$ & $-1927.24$ & $-1928.33$ & $-1930.40$ & $-1931.25$ \\
        $\log_{10}BF$ & - & $0$ & $-0.43$ & $-1.52$ & $-3.59$ & $-4.44$ \\
        \hline
        $f_1$ & U$(0, 1)$ & $0.54^{+0.35}_{-0.43}$ & $0.53^{+0.33}_{-0.41}$ & $0.58^{+0.32}_{-0.35}$ & $0.38^{+0.42}_{-0.36}$ & $0.62^{+0.27}_{-0.26}$ \\
        $f_2$ & U$(0, 1-f_1)$ & \colorbox{gray!25}{$0.46^{+0.43}_{-0.35}$} & \colorbox{gray!25}{$0.47^{+0.41}_{-0.33}$} & \colorbox{gray!25}{$0.42^{+0.35}_{-0.32}$} & $0.55^{+0.29}_{-0.42}$ & $0.36^{+0.27}_{-0.27}$ \\
        $f_3$ & U$(0, 1-f_1-f_2)$ & - & - & - & \colorbox{gray!25}{$0.06^{+0.19}_{-0.05}$} & \colorbox{gray!25}{$0.02^{+0.05}_{-0.02}$} \\
        \hline
        $\beta_1$ & U$(-2, 7)$ & $1.80^{+1.75}_{-1.22}$ & $1.29^{+2.21}_{-1.15}$ & $1.06^{+1.11}_{-0.96}$ & $2.56^{+2.94}_{-3.78}$ & $1.40^{+1.41}_{-1.01}$ \\
        $\beta_2$ & U$(-2, 7)$ & $2.81^{+3.31}_{-3.84}$ & $1.47^{+4.44}_{-2.96}$ & $3.15^{+3.39}_{-3.95}$ & $2.75^{+3.48}_{-2.79}$ & $2.97^{+3.46}_{-3.88}$ \\
        $\beta_3$ & U$(-2, 7)$ & - & - & - & $0.38^{+5.57}_{-2.17}$ & $1.56^{+4.47}_{-3.20}$ \\
        $\xi_2$ & U$(0, 1)$ & $0.46^{+0.42}_{-0.43}$ & $0.48^{+0.44}_{-0.42}$ & $0.45^{+0.48}_{-0.39}$ & $0.68^{+0.27}_{-0.41}$ & $0.42^{+0.50}_{-0.36}$ \\
        $\xi_3$ & U$(0, 1)$ & - & - & - & $0.59^{+0.35}_{-0.35}$ & $0.46^{+0.47}_{-0.41}$ \\
        $\mu_{q,2}$ & U$(0, 1)$ & $0.56^{+0.37}_{-0.40}$ & $0.55^{+0.38}_{-0.48}$ & $0.52^{+0.41}_{-0.46}$ & $0.66^{+0.30}_{-0.40}$ & $0.60^{+0.35}_{-0.50}$ \\
        $\mu_{q,3}$ & U$(0, 1)$ & - & - & - & $0.51^{+0.36}_{-0.35}$ & $0.43^{+0.47}_{-0.37}$ \\
        $\sigma_{q,2}$ & U$(0.01, 1)$ & $0.58^{+0.35}_{-0.37}$ & $0.54^{+0.40}_{-0.44}$ & $0.50^{+0.44}_{-0.41}$ & $0.56^{+0.34}_{-0.41}$ & $0.51^{+0.43}_{-0.43}$ \\
        $\sigma_{q,3}$ & U$(0.01, 1)$ & - & - & - & $0.78^{+0.20}_{-0.30}$ & $0.60^{+0.36}_{-0.48}$ \\
        $m_\text{gap}$ ($M_\odot$) & U$(20, 80)$ & $66.31^{+11.10}_{-11.98}$ & $65.36^{+12.19}_{-16.13}$ & $76.88^{+2.75}_{-4.23}$ & $61.31 ^{+12.16}_{-38.34}$ & $76.75^{+2.87}_{-4.28}$ \\
        \hline
        $A_1$ & U$(0, 1)$ & $0.09^{+0.63}_{-0.09}$ & $0.96^{+0.04}_{-0.56}$ & $0.95^{+0.05}_{-0.10}$ & $0.93^{+0.06}_{-0.51}$ & $0.95^{+0.05}_{-0.10}$ \\
        $A_2$ & U$(0, 1)$ & $0.49^{+0.33}_{-0.37}$ & $0.21^{+0.71}_{-0.19}$ & $0.35^{+0.54}_{-0.31}$ & $0.29^{+0.33}_{-0.26}$ & $0.41^{+0.47}_{-0.36}$ \\
        $A_3$ & U$(0, 1)$ & - & - & - & $0.59^{+0.40}_{-0.29}$ & $0.41^{+0.48}_{-0.35}$ \\
        $n_1$ & U$(0, 3)$ & $1.61^{+1.15}_{-1.47}$ & $0.37^{+1.20}_{-0.31}$ & $0.41^{+0.55}_{-0.35}$ & $0.57^{+1.81}_{-0.52}$ & $0.36^{+0.54}_{-0.31}$ \\
        $n_2$ & U$(0, 3)$ & $1.84^{+1.04}_{-1.63}$ & $1.37^{+1.36}_{-1.20}$ & $1.62^{+1.19}_{-1.43}$ & $1.10^{+1.39}_{-0.94}$ & $1.70^{+1.17}_{-1.39}$ \\
        $n_3$ & U$(0, 3)$ & - & - & - & $1.45^{+1.02}_{-1.21}$ & $1.66^{+1.19}_{-1.44}$ \\
        $\mu^{\chieff}_1$ & U$(0, 1)$ & $0.03^{+0.02}_{-0.02}$ & $0.03^{+0.02}_{-0.46}$ & $0.03^{+0.02}_{-0.02}$ & $0.03^{+0.03}_{-0.74}$ & $0.03^{+0.02}_{-0.02}$ \\
        $\mu^{\chieff}_2$ & U$(0, 1)$ & $-0.07^{+0.18}_{-0.26}$ & $-0.09^{+0.62}_{-0.74}$ & $0.03^{+0.53}_{-0.80}$ & $-0.18^{+0.65}_{-0.59}$ & $0.03^{+0.61}_{-0.88}$ \\
        $\mu^{\chieff}_3$ & U$(0, 1)$ & - & - & - & $-0.04^{+0.08}_{-0.08}$ & $0.56^{+0.38}_{-0.47}$ \\
        $\sigma^{\chieff}_1$ & U$(0.01, 1)$ & $0.35^{+0.50}_{-0.31}$ & $0.06^{+0.42}_{-0.02}$ & $0.06^{+0.02}_{-0.01}$ & $0.06^{+0.02}_{-0.01}$ & $0.06^{+0.02}_{-0.01}$ \\
        $\sigma^{\chieff}_2$ & U$(0.01, 1)$ & $0.26^{+0.43}_{-0.19}$ & $0.47^{+0.41}_{-0.41}$ & $0.33^{+0.53}_{-0.27}$ & $0.48^{+0.46}_{-0.41}$ & $0.40^{+0.46}_{-0.34}$ \\
        $\sigma^{\chieff}_3$ & U$(0.01, 1)$ & - & - & - & $0.06^{+0.11}_{-0.04}$ & $0.56^{+0.38}_{-0.47}$ \\
        $\chi_1^\text{min}$ & U$(-1, \chi_1^\text{max})$ & $-0.09^{+0.05}_{-0.06}$ & - & - & - & - \\
        $\chi_1^\text{max}$ & U$(-1, 1)$ & $0.14^{+0.03}_{-0.03}$ & - & - & - & - \\
        $\chi_2^\text{min}$ & U$(-1, \chi_2^\text{max})$ & $-0.19^{+0.47}_{-0.79}$ & - & - & - & - \\
        $\chi_2^\text{max}$ & U$(-1, 1)$ & $0.50^{+0.14}_{-1.47}$ & - & - & - & - \\
        $\kappa_1$ & U$(-10, 10)$ & $2.61^{+1.36}_{-1.00}$ & $2.67^{+4.32}_{-0.87}$ & $2.7^{+0.94}_{-1.03}$ & - & - \\
        $\kappa_2$ & U$(-10, 10)$ & $4.96^{+1.81}_{-2.05}$ & $5.72^{+2.18}_{-2.51}$ & $6.44^{+2.33}_{-2.23}$ & $7.48^{+1.48}_{-1.99}$ & $6.19^{+3.06}_{-2.76}$ \\
        $d$ & U$(-5, 10)$ & - & - & - & $6.42^{+3.01}_{-3.84}$ & $6.80^{+2.89}_{-3.88}$ \\
        $t_d^{min}$ (Gyr) & U$(0.01, 10)$ & - & - & - & $0.86^{+0.90}_{-0.60}$ & $1.14^{+0.99}_{-0.83}$ \\
        \hline
    \end{tabular}
    \caption{Priors used for the mass ratio, spin and redshift model hyperparameters and mixing fractions for the two- and three-channel models as well as corresponding posterior medians and 90\% credible interval errors. Estimates for mixing fractions highlighted in grey are not inferred directly but obtained from the constraint $\sum f_i =1$.}
    \label{tab:q_spins_z_hyperpriors}
\end{table*}

\newpage

\bibliography{references}{}
\bibliographystyle{aasjournalv7}

\end{document}